# On the Phase Structure of the Schwinger Model with Wilson Fermions *

H. Gausterer and C.B. Lang

Institut für Theoretische Physik,
Universität Graz, A-8010 Graz, AUSTRIA

We study the phase structure of the massive one flavour lattice Schwinger model on the basis of the finite size scaling behaviour of the partition function zeroes. At $\beta = 0$ we observe and discuss a possible discrepancy with results obtained by a different method.

## 1. INTRODUCTION

Since its introduction by Schwinger in 1962 [1], two dimensional QED has been continually examined and has attracted both high energy and mathematical physicists. The model shows important aspects of realistic models, is exactly solvable and the Wightman axioms hold for the bosonized fields [2]. Among the properties are breaking of chiral symmetry (mass generation) based on confinement properties, charge shielding and the topologically interesting fact of a continuum of degenerated vacua labeled by an angle $\theta$. Certain properties extend to the massive version of the Schwinger model, where chiral symmetry is explicitly broken [3]. The lattice version of the Schwinger model has served extensively as a laboratory to test fermionic algorithms, mainly for Kogut-Susskind fermions. Its formal continuum limit from a constructive field theory point of view was examined in [4].

In particular the Schwinger model with fermions on the Wilson representation may teach us how to approach continuum physics due to its analogy to lattice $QCD$ with Wilson fermions. Both models *are believed* to have similar zero temperature phase structure: For any gauge coupling $\beta$ there exists a critical hopping parameter $\kappa_c(\beta)$. Thus continuum physics could be approached along this line which ends at $\kappa_c(\infty) = 1/(2d)$, a critical point with a second order transition.

Up to now it is (from a rigorous point of view) unproven and thus a matter of believe how far this line really extends into the $\beta$–$\kappa$ plane. It is assumed and supported by strong coupling data, that there is a critical point at $\beta = \infty$ with $0 < \kappa_c < \infty$. In the strong coupling limit expectation values of operators are expanded around $(\kappa, \beta) = 0$.

$$\int d\mu(U) O(U) \det M(\kappa, U) e^{-S_G(\beta,U)} = \sum_{n,m} c_{n,m} \kappa^n \beta^m \ . \tag{1}$$

Here $M$ denotes the lattice Dirac operator. This expansion is known to converge at least in a nonvanishing domain around the expansion point, i.e. $|\kappa| < \rho_\kappa(\beta)$ and $\beta < \rho_\beta$ [5] and has in that range a finite mass gap. Usually $\rho_\kappa(\beta)$ is then identified with a critical value $\kappa_c(\beta) \in \mathbb{R}^+$. This, however, is not necessarily true, since the singularity could occur for a complex $\kappa$ on the circle $|\kappa| = \rho_\kappa(\beta)$.

For the Schwinger model it can be shown that $\rho_\kappa(0) \leq \frac{1}{2}$ [6]. Further checks, e.g. numerical simulations, have to be done to verify the assumption $\kappa_c = \rho_\kappa$. Typically one would calculate the chiral condensate $\langle\bar{\psi}\psi\rangle$, but for Wilson fermions this is not an order parameter and no simple parameter is known. Another standard possibility is to determine the correlation length (inverse mass gap) and look for divergences in the plane of coupling constants. Doing this in an analytic calculation (hopping expansion) the necessary approximations are too rough for a trustworthy verification whether there is a critical point.

A direct numerical simulation of the one flavour Schwinger model for general $\beta$ is not possible with present means since $\det M(\kappa, U)$ as well as quadratic forms $(\phi, M^{-1}(\kappa, U)\phi)$ are not positive definite for certain gauge field configurations. In this case Metropolis type algorithms will fail. It has been shown in [7] that the one flavour Schwinger model at $\beta = 0$ can be mapped onto a "seven-vertex" model (the eight-vertex model for a particular choice of weight parameters). In the limit $\kappa \to \infty$ that model becomes the critical six-vertex model [8]. Therefore the Schwinger model has at least one critical point at $\beta = 0$ and $\kappa_c(0) = \infty$, far beyond the range of the cluster expansion. This map further offers the opportunity for a direct simulation of the one flavour Schwinger model at least at $\beta = 0$ and allows for a test of strong coupling assumptions on criticality. The model was simulated in [6]; with the available statistics the results did not show a clear or conclusive signal for a critical behaviour for $\kappa_c(0) \leq \frac{1}{2}$. Rather they indicate that there might be no critical point but just a cross-over like behaviour. So either there is indeed no critical point or the available statistics is by far not sufficient.

*Supported by Fonds zur Förderung der Wissenschaftlichen Forschung in Österreich, project P7849.

## 2. PARTITION FUNCTION ZEROS

An alternative approach to investigate the phase structure is to determine the partition function of the Schwinger model for fixed $\beta$ and $\kappa \in \mathbb{C}$ and to study the finite size scaling behaviour of the closest Lee–Yang zeroes [9,10]. The partition function is given by

$$\begin{aligned} Z_{S,\Lambda}(\kappa) &= \int d\mu(U) \det M_\Lambda(\kappa, U) e^{-S_G(\beta,U)} \\ &= Z_{G,\Lambda} \langle \det M_\Lambda(\kappa, U) \rangle \ , \end{aligned} \quad (2)$$

which is a polynomial of degree $2|\Lambda|$ in $\kappa$. For a finite system the zeroes in $Z_{S,\Lambda}(\kappa)$ are strictly complex and the partition function and the free energy are analytic in a non-vanishing neighbourhood of the real axis. As criticality is approached (in the thermodynamic limit and for $\kappa \to \kappa_c$) the Lee–Yang zeroes pinch the real $\kappa$ axis, inducing a phase transition. The finite size scaling behaviour of these zeroes, in particular those closest to the real axis, is related to the thermodynamic critical exponents [11,10].

There are three ways to determine $\langle D \rangle \equiv \langle \det M_\Lambda(\kappa, U) \rangle$ for complex $\kappa$.

- One obtains the coefficients of the polynomial expressed in terms of moments of $M_\Lambda^{-1}$, by a direct numerical simulation. However, since coefficients may not decay sufficiently fast with $n$, we may have to find all coefficients. This approach is closest related to the, elsewhere very successful, method of analytic extrapolation via histograms [10].

- One may calculate values $\langle D \rangle$ with sufficient precision for various real $\kappa$, perform a polynomial fit and use this fit to analytically continue to complex $\kappa$. In a test of this method for the analytically solvable free fermion theory it turned out to be very unstable and could not be used to identify the closest zeroes.

- Finally the hard but direct way: Calculate $\langle D \rangle$ for $\kappa \in \mathbb{C}$ and scan certain regions of the complex plane for the closest zeroes. This method is applicable for small lattices only, since one has to determine the determinant for each gauge field configuration explicitly worked. On the other hand it does not involve any extrapolation and works for arbitrary number of flavours.

We used the last mentioned method in our analysis; since it is very CPU-time consuming we were restricted to relatively small lattices of the size $|\Lambda| = 2^2$, $4^2$, $8^2$. Due to the smallness of the lattices a conclusive statement on the existence of a critical point may not possible. However, comparing the finites size scaling behaviour of the zeroes with that of the analytically known free fermion theory does give a hint what to expect.

In fig. 1 we show the positions of the closest zeroes for the three values of $\beta = 0$, 1 and 5. A direct comparison with the values at $\beta = \infty$ is not possible, since there the absolute positions depend on the boundary conditions. (For finite $\beta$ and $U(1)$ gauge group the b. c. for the fermions are irrelevant and averaged out anyhow). In fig. 2 we compare the scaling behaviour of the imaginary part of the closest zeroes with that for the free fermion theory. We find no significant

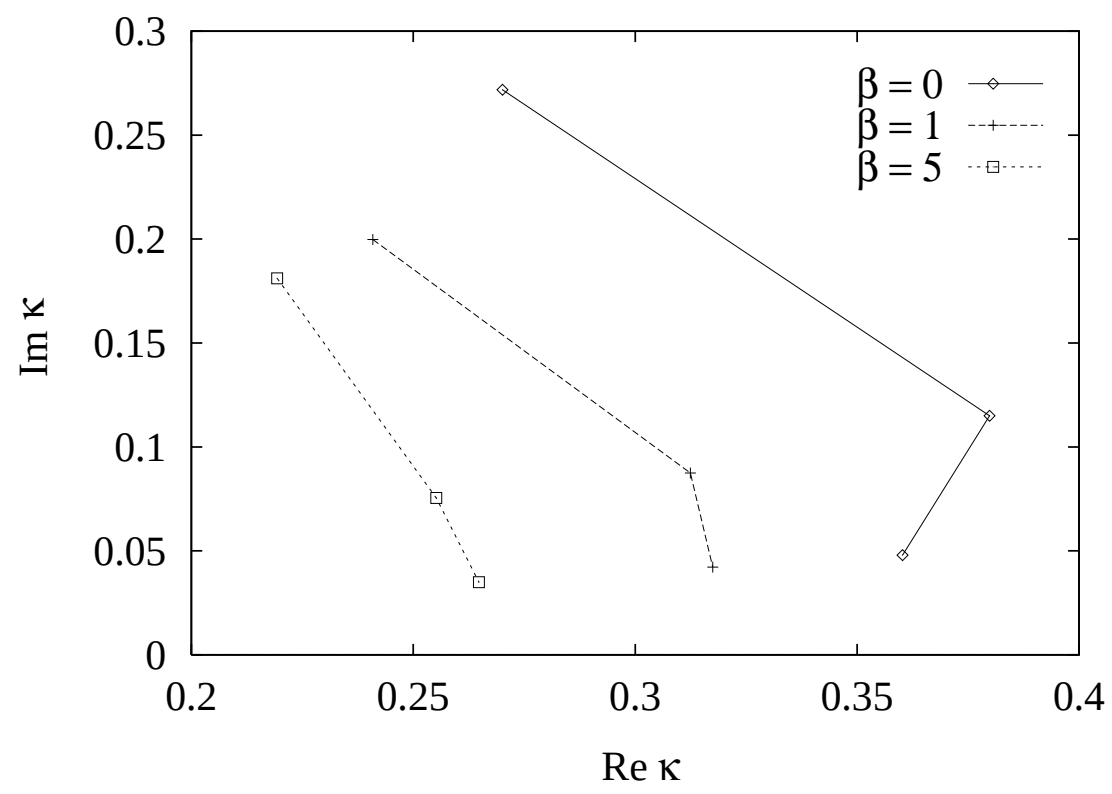

Figure 1. Position of the closest zero in the complex $\kappa$ plane for lattice sizes $L = 2$, 4, and 8 and gauge coupling $\beta = 0$, 1, and 5.

difference and observe a scaling compatible with $\mathrm{Im}(\kappa_0) \propto 1/L$.

All data are consistent with the assumption that there is a critical $\kappa_c$ for these gauge couplings. This is particularly interesting for $\beta = 0$ where the eight-vertex data [6] suggested that there is no critical point at finite $\kappa(0)$. On the other hand, much larger lattices were studied there. Further studies with enhanced statistics may clarify this discrepancy.

The phase structure extrapolated from the scaling behaviour of the zeroes is thus in agreement with the assumption that there is a critical line starting at $\kappa_c(\infty) = \frac{1}{4}$ continuing through all values of $\beta$ and ending at a $\kappa_c(0) \leq \frac{1}{2}$.

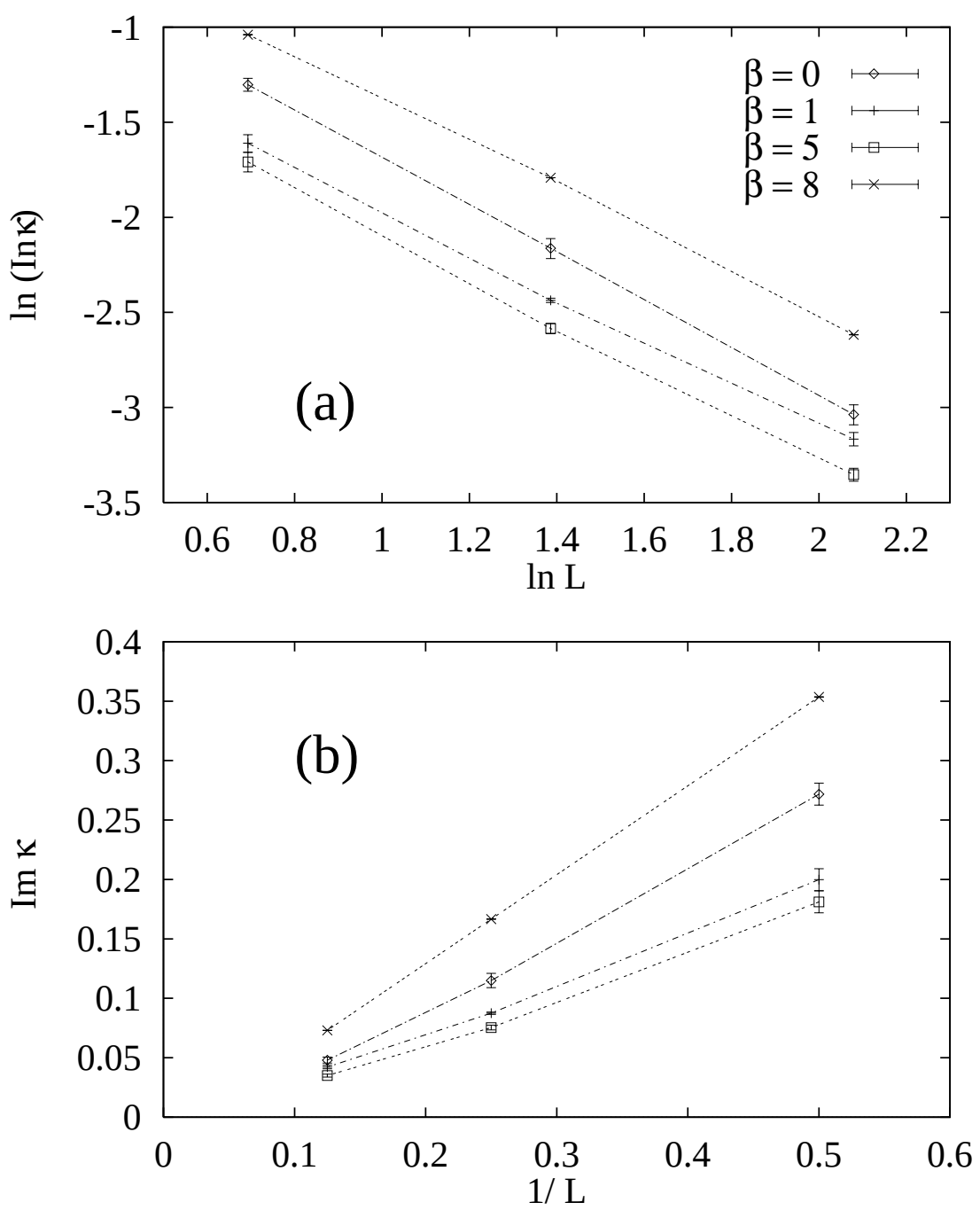

Figure 2. The imaginary part of the closest Lee–Yang zero $\mathrm{Im}(\kappa_0)$ as function of the lattice size $L$. (a) A ln-ln plot, comparing values at $\beta = 0$, 1, and 5 with those at $\infty$. (b) Scaling plot vs. $1/L$, the scaling behaviour expected for $\beta = \infty$.